\documentclass[%
 amsmath,amssymb,
 aps,
]{revtex4-1}

\usepackage{float} 
\usepackage{graphicx}
\usepackage{dcolumn}
\usepackage{bm}
\usepackage[english]{babel}
\usepackage{color}
\usepackage{hyperref}
\hypersetup{
    colorlinks=true,
    linkcolor=blue,
    filecolor=blue,      
    urlcolor=blue,
    citecolor=blue
}

\begin{document}


\title{Analytical approximations for the dispersion of electromagnetic modes in slabs of biaxial crystals}

\author{Gonzalo \'Alvarez-P\'erez\textsuperscript{1,2}}
\author{Kirill V. Voronin\textsuperscript{3}}
\author{Valentyn S. Volkov\textsuperscript{3}}
\author{Pablo Alonso-Gonz\'alez\textsuperscript{1,2}}
\altaffiliation{Corresponding author: pabloalonso@uniovi.es}
\author{Alexey Y. Nikitin\textsuperscript{4,5,3}}
\altaffiliation{Corresponding author: alexey@dipc.org}
\affiliation{\vspace{10pt}
\textsuperscript{1}Department of Physics, University of Oviedo, Oviedo 33006, Spain.\\
\textsuperscript{2}Center of Research on Nanomaterials and Nanotechnology, CINN (CSIC--Universidad de Oviedo), El Entrego 33940, Spain.\\
\textsuperscript{3}Center for Photonics and 2D Materials, Moscow Institute of Physics and Technology, Dolgoprudny 141700, Russia.\\
\textsuperscript{4}Donostia International Physics Center (DIPC), Donostia-San Sebast\'an 20018, Spain.\\
\textsuperscript{5}IKERBASQUE, Basque Foundation for Science, Bilbao 48013, Spain.
}

\date{\today}

\begin{abstract}

Anisotropic crystals have recently attracted considerable attention because of their ability to support polaritons with a variety of unique properties, such as hyperbolic dispersion, negative phase velocity, or extreme confinement. Particularly, the biaxial crystal $\alpha$-MoO\textsubscript{3} has been demonstrated to support phonon polaritons, light coupled to lattice vibrations, with in-plane anisotropic propagation and unusually long lifetime. However, the lack of theoretical studies on electromagnetic modes in biaxial crystal slabs impedes a complete interpretation of the experimental data, as well as an efficient design of nanostructures supporting such highly anisotropic polaritons. Here, we derive the dispersion relation of electromagnetic modes in biaxial slabs surrounded by semi-infinite isotropic dielectric half-spaces with arbitrary dielectric permittivities. Apart from a general dispersion relation, we provide very simple analytical expressions in typical experiments in nano-optics: the limits of short polaritonic wavelength and/or very thin slabs. The results of our study will allow for an in-depth analysis of anisotropic polaritons in novel biaxial van der Waals materials.
\end{abstract}

\maketitle


\section{\label{sec:Introduction}Introduction}

Anisotropic media have been a subject of fundamental
and applied research in optics for several centuries since the
earliest Bartholinus’ studies. Particularly, birefringence, responsible for the double refraction inside anisotropic crystals, is widely used nowadays in daily-life applications requiring polarization filtering as, for instance, sun glasses, liquid crystal displays, or scanning laser polarimetry (for monitoring glaucoma) \cite{Kliger90}. In recent decades, the scientific interest to anisotropic optical phenomena has dramatically increased due to the design and fabrication of novel artificial materials (metamaterials) with a tailored optical response. Striking examples of the latter are photonic and plasmonic crystals \cite{Yablonovitch87, Barnes03, Pendry04} and metasurfaces \cite{Yu14, Kildishev13}, showing spectacular phenomena such as negative refraction \cite{Cubukcu03}, slow light \cite{Baba08}, and superlensing \cite{Pendry00, Taubner06}, among others. Apart from these anisotropic artificial materials, a few years ago the concept of ''atomic-scale'' engineering with naturally anisotropic van der Waals (vdW) materials \cite{Geim13} was suggested, adding more scientific interest to this field. Currently, presenting one of the main strategies in low-dimensional optoelectronics, this concept has induced an intensive study of highly-confined anisotropic polaritons supported by vdW slabs and heterostructures \cite{Basov16, Low17}. The possibility of visualizing these polaritons in thin slabs of vdW crystals with the use of near-field microscopy \cite{Dai14, Li18, Ma18, Dai15, Zheng18} stimulates more and more experimental and theoretical studies in this direction.

From a theoretical point of view, in bulk uniaxial crystals (such as $h$-BN, SiC or layered metamaterials), characterized by two refractive indices, electromagnetic eigenmodes present ordinary and extraordinary waves. In many cases, the propagation of light along the boundaries of uniaxial crystals and inside the slabs can be straightforwardly analyzed analytically \cite{Dyakonov88, Takayama08, Dai14}. In stark contrast, biaxial crystals (such as $\alpha$-MoO\textsubscript{3} or V\textsubscript{2}O\textsubscript{5}) are characterized by three refractive indices and both electromagnetic eigenmodes are extraordinary. As a result, the understanding and analytical treatment of electromagnetic phenomena in biaxial media is significantly more complex than in the uniaxial case. In this context, a very recent study has reported on a rigorous analytical solution for the dispersion of surface waves on the boundaries of biaxial crystals \cite{Narimanov18}; however, up to now, studies on the electromagnetic modes in biaxial slabs have been mainly the subject of a numerical analysis \cite{Maldonado96,Kharusi74, Jovanovic01} or some particular configurations, such as grounded crystal slabs with modes fixed propagation directions \cite{Chen18}.

In this work, organized in a tutorial style, we present a detailed derivation of the dispersion relation of electromagnetic modes in a biaxial slab of a finite thickness (with arbitrary dielectric tensor), surrounded by two semi-infinite isotropic media with arbitrary dielectric permittivities. We assume that one of the principle crystal axes is perpendicular to the faces of the slab, while the mode propagates along the crystal slab at an arbitrary angle with respect to the other principal axes, which lays in a plane parallel to the faces of the slab. We show that our general dispersion relation successfully reduces to the known limiting cases, such as the case of a uniaxial slab or a semi-infinite crystal, among others. We manage to reduce the general dispersion relation to simple analytical expressions for short wavelength of the modes and small slab thicknesses, which are currently of great interest for the study of anisotropic polaritons in vdW slabs. To demonstrate the validity of our analytical approximations, we compare them to full-wave simulations, finding an excellent agreement.

\section{\label{sec:Frensel}Infinite biaxial crystal}

\begin{figure}[H]
\centering
\includegraphics[width=8.5cm]{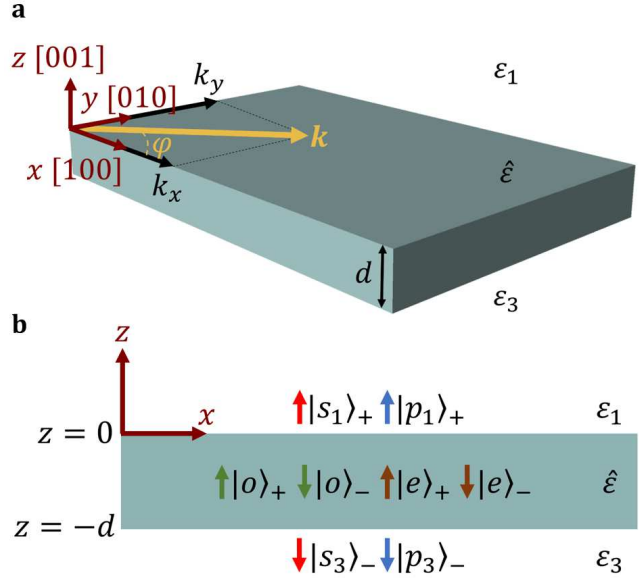}
\caption{\label{Fig1} Schematics of the biaxial slab. One of the main crystal axes $[0,0,1]$ is perpendicular to the faces of the slab and coincides with the $z$-axis, while the axes $[1,0,0]$ and $[0,1,0]$ belong to a plane parallel to the faces of the slab and are directed along the coordinate axes $x$ and $y$, respectively. The mode propagates at an arbitrary angle $\varphi$ with respect to the $x$ axis.}
\label{SNOM}
\end{figure}

Let us consider an infinite, nonmagnetic biaxial medium with dielectric permittivity tensor $\hat{\varepsilon}$. The coordinate system $\lbrace x,y,z \rbrace$ is chosen is such a way that $\hat{\varepsilon}$ is diagonal [see Fig. \ref{Fig1}(a)], so that
\begin{equation}
\hat{\varepsilon}=\left( \begin{array}{ccc}
     \varepsilon_x & 0 & 0 \\
     0 & \varepsilon_y & 0 \\
     0 & 0 & \varepsilon_z 
\end{array} \right).
\label{S2Eps2}
\end{equation}

To accurately decompose the electromagnetic fields in the biaxial medium, we need to define appropriate basis vectors. To that end, we follow a standard procedure, as for example in Ref. \cite{Narimanov18}. Namely, we represent the electric and magnetic fields in the biaxial medium in the form of plane waves:
\begin{equation}
    \textbf{E}=E_0 \textbf{e}\: e^{i\textbf{k} \textbf{r}-i\omega t}, \qquad \qquad 
    \textbf{H}=H_0 \textbf{h}\: e^{i\textbf{k} \textbf{r}-i\omega t},
\label{S2Fields}
\end{equation}

\noindent where $\textbf{e}$ and $\textbf{h}$ are unknown dimensionless field basis vectors, $E_0$ and $H_0$ are arbitrary field amplitude coefficients, $\omega$ is the angular frequency, $\textbf{k}$ is the wave vector, and $\textbf{r}$ is the radius vector.

From Maxwell's equations ($\nabla \times \textbf{E}= -\frac{1}{c} \frac{\partial\textbf{H}}{\partial t}$ and $\nabla \times \textbf{H}= \frac{1}{c} \frac{\partial\textbf{D}}{\partial t}$), and substituting the magnetic field, $\textbf{H}$, we obtain a vectorial equation for the electric fields, $\textbf{E}$:
\begin{equation}
\frac{\omega^2}{c^2}\hat{\varepsilon} \textbf{E}=\nabla\left(\nabla \textbf{E}\right)-\Delta \textbf{E}.
\label{S2Max}
\end{equation}

Substituting $\textbf{E}$ from Eq. (\ref{S2Fields}) into Eq. (\ref{S2Max}), we obtain a linear homogeneous system of equations for the three components of the unknown basis vector, $\textbf{e}$:
\begin{equation}
\mathcal{M}\;\textbf{e}=\left( \begin{array} {ccc} 
\Delta_x & q_{x}q_{y} & \pm i q_{x}q_{z}\\
q_{x}q_{y} & \Delta_y & \pm i q_{y}q_{z}\\
\pm i q_{x}q_{z} & \pm i q_{y}q_{z} & \Delta_z\end{array} \right) 
\left( \begin{array} {ccc} 
e_x\\
e_y\\
e_z\end{array} \right)=0,
\label{S2Syst}
\end{equation}

\noindent where $k_0=\omega / c$ is the free space wave vector, $q_{x,y}=k_{x,y}/k_0$ are the in-plane components of the normalized wave vector, and $q_{z}$ is the out-of-plane component of the normalized wave vector, so that $k_z=\pm i q_{z} k_0$. The $+$($-$) sign must be taken for the wave propagating along (opposite to) the z--axis, while $\Delta_i$ are defined as
\begin{equation}
\begin{split}
\Delta_x = \varepsilon_x - q_y^2 + q_z^2, \\
\Delta_y = \varepsilon_y - q_x^2 + q_z^2, \\
\Delta_z = \varepsilon_z - q_x^2 - q_y^2.
\end{split}
\label{S2Delta}
\end{equation}

Our choice of the dependence of the basis vectors upon the coordinates (the propagation along $z$ is treated differently from the propagation along $x$ and $y$) is dictated by the geometry of the problem (see Fig. \ref{Fig1}), consistently with the standard waveguide theory.

The system (\ref{S2Syst}) has nontrivial solutions only when $\text{det}\left(\mathcal{M}\right)=0$, that gives the well-known Fresnel's equation for biaxial media \cite{Landau60,Born99}:
\begin{equation}
q_z^2 \left[q_z^2 \varepsilon_z + \varepsilon_z (\varepsilon_x + \varepsilon_y) - q_x^2 (\varepsilon_x + \varepsilon_z) - q_y^2 (\varepsilon_y + \varepsilon_z)\right] + (\varepsilon_z - q_x^2- q_y^2)(\varepsilon_x \varepsilon_y - q_x^2 \varepsilon_x - q_y^2 \varepsilon_y)=0.
\label{S2FrEq}
\end{equation}

It is the quadratic equation in terms of the squared z-component of the wave vector, $q_z$.  Its solutions $q_{ez}$ and $q_{oz}$ read as
\begin{equation}
q_{o,ez}^{2}=\frac{1}{2}\left\lbrace \frac{\varepsilon_{x} + \varepsilon_{z}}{\varepsilon_{z}} q_{x}^{2} + \frac{\varepsilon_{y} + \varepsilon_{z}}{\varepsilon_{z}} q_{y}^{2} - (\varepsilon_{x} + \varepsilon_{y})\right\rbrace \pm \frac{1}{2} \sqrt{D},
\label{S2WV}
\end{equation}

\noindent with $D$ being the discriminant
\begin{equation}
D=\left(\varepsilon_{x}-\varepsilon_{y}+ \frac{\varepsilon_{z}-\varepsilon_{x}}{\varepsilon_{z}} q_{x}^2-\frac{\varepsilon_{z}- \varepsilon_{y}}{\varepsilon_{z}} q_{y}^{2}\right)^{2} + 4\frac{(\varepsilon_{z} - \varepsilon_{x})(\varepsilon_{z}- \varepsilon_{y})}{\varepsilon_{z}^{2}} q_{x}^{2} q_{y}^{2}.
\label{S2Det}
\end{equation}

In Eq. (\ref{S2WV}), the sign "$+$" and "$-$" correspond to the labels "$o$", and "$e$", respectively. Substituting Eq. (\ref{S2WV}) into the system (\ref{S2Syst}), we find all three components of the two eigenvectors $\textbf{e}$. Since the system (\ref{S2Syst}) is homogeneous, one of the components of the eigenvectors must be fixed. Without loss of generality, fixing the $y$-component to $e_y=q_x$ for the root "$o$" and to $e_y=q_y$  for the root "$e$", we find
\begin{align}
    \textbf{e}_{o} = \frac{1}{q}\left(\begin{array}{c}
         -q_y (1 - \Delta_1 \Delta_z) \\
         q_x \\
         \mp i  q_x q_y q_{oz} \Delta_1
    \end{array}  \right), \qquad \qquad 
    \textbf{e}_{e} = \frac{1}{q}\left(\begin{array}{c}
         q_x\frac{\Delta_2 - q_y^2}{\Delta_x^e} \\
         q_y \\
         \frac{\Delta_2}{\mp i  q_{ez}}
    \end{array}  \right),
\label{S2Basis}    
\end{align}
where the factor $1/q$ ($q$ being the normalized in-plane wave vector, $q^2= \frac{k_x^2 + k_y^2}{k_0^2}$) stands for the normalization, and
\begin{equation}
    \Delta_1 = \frac{\Delta_x^o - q_x^2}{\Delta_z \Delta_x^o + q_x^2 q_{oz}^2}, \quad \Delta_2 = \frac{\Delta_x^e\Delta_y^e - q_x^2 q_y^2}{\Delta_x^e - q_x^2}.
\label{const_delta}
\end{equation}
From Eq. (\ref{S2FrEq}) we can easily find the asymptotes of the isofrequency curves for large $q_{x,y}$. Tending both $q_{x}$ and $q_{y}$ to infinity, and setting $q_z=0$, we find
\begin{equation}
\frac{q_x}{q_y} = \sqrt{-\frac{\varepsilon_y}{\varepsilon_x}}.
\label{asF}
\end{equation}
In thin vdW slabs these asymptotes yield the direction of the propagation of the polaritonic "rays", excited by localized sources \cite{Dai15, Ma18, Li18}.
In the particular case of a uniaxial crystal (with the axis $C$ pointing parallel to the $z$-axis, $C\| Oz$),  $\varepsilon_x= \varepsilon_y= \varepsilon_\bot$ and $\varepsilon_z= \varepsilon_\parallel$, the derived basis vectors \eqref{S2Basis} can be straightforwardly transformed to the basis vectors for the ordinary and extraordinary waves. Taking into account that the $z$-components of the wave vectors (\ref{S2WV}) are reduced to the well-known expressions for the ordinary and extraordinary waves
\begin{equation}
q_{oz}^2 = q^2 - \varepsilon_\bot,\quad
q_{ez}^2 = \frac{\varepsilon_\bot}{\varepsilon_\parallel} q^2 - \varepsilon_\bot,
\label{oe}
\end{equation}

\noindent we find that $\Delta_1=0$, $\Delta_2= \varepsilon_\bot + q_{ez}^2$ and obtain the basis vectors
\begin{align}
\begin{split}
    \textbf{e}_o = \frac{1}{q} \left(\begin{array}{c}
         -q_y \\
         q_x \\
         0
    \end{array}  \right), \qquad \qquad
    \textbf{e}_e = \frac{1}{q} \left(\begin{array}{c}
         q_x \\
         q_y \\
         \frac{\varepsilon_\bot + q_{ez}^2}{\mp i q_{ez}}
    \end{array}  \right).
    \end{split}
\label{S2Uniax}    
\end{align}
In case of an isotropic medium,  $\varepsilon_\bot= \varepsilon_\parallel=\varepsilon$, the $z$-components of the wave vectors degenerate $q_{oz}^2=q_{ez}^2=q_z^2 = q^2 - \varepsilon$ and the basis vectors \eqref{S2Basis} reduce to the ones for the  $s$- and $p$-polarized waves ($\textbf{e}_o \rightarrow \textbf{e}_s$ and $\textbf{e}_e \rightarrow \textbf{e}_p$):
\begin{align}
\begin{split}
   \textbf{e}_s = \frac{1}{q}\left(\begin{array}{c}
         -q_y \\
         q_x \\
         0
    \end{array}  \right), \qquad \qquad
    \textbf{e}_p = \frac{1}{q}\left(\begin{array}{c}
         q_x \\
         q_y \\
         \frac{q^2}{\mp i q_{z}}
    \end{array}  \right).
\label{S2Isotr}    
\end{split}
\end{align}

\section{\label{sec:Biaxial}Biaxial slab of a finite thickness}

Here, we derive the dispersion relation for polaritons in a biaxial slab of thickness $d$ and permittivity $\hat{\varepsilon}$, occupying the region $0>z>-d$ between two dielectric half-spaces with permittivities $\varepsilon_1$ (region "1", $z>0$) and $\varepsilon_3$ (region "3", $z<-d$).

\subsection{\label{sec:level2}General form of the dispersion relation}
Let us first represent the electric fields above ($z>0$) and below ($z<-d$) the slab, in the isotropic media "1" and "3", respectively. In these regions we can take the fields in the form of the $s$- and $p$-polarized plane waves.  For compactness, from now on, we will use Dirac notation, in which the $s$- and $p$-polarization basis vectors read

\begin{equation}
\begin{split}
    \vert s_{1,3} \rangle_{\pm} = \frac{1}{q}\left( \begin{array}{c}
     -q_y \\
     q_x \\
     0
\end{array} \right) e^{ik_x x + i k_y y}, \qquad \qquad
\vert p_{1,3} \rangle_{\pm} = \frac{1}{q}\left( \begin{array}{c}
     q_x \\
     q_y \\
     \frac{q^2}{\mp iq_{1,3z}}
\end{array} \right) e^{ik_x x + i k_y y},
\end{split}
\label{S31}
\end{equation}

\noindent where $q_{1,3z}=\sqrt{q_x^2 + q_y^2 - \varepsilon_{1,3}} >0$ is the out-of-plane component of the normalized wave vector. Here and in the definition of the $p$-polarization basis vector, $\vert p_{1,3} \rangle_{\pm}$, the $+$($-$) sign should be taken for the wave propagating along (opposite to) the z--axis, while in case of the $s$-polarization $\vert s_{1,3} \rangle_{+}$ and $\vert s_{1,3} \rangle_{-}$ are degenerated.  For convenience, we also introduce the in-plane subvectors of the vectors given by Eq. (\ref{S31}):
\begin{equation}
\begin{split}
    \vert s \rangle = \frac{1}{q}\left( \begin{array}{c}
     -q_y \\
     q_x 
\end{array} \right) e^{ik_x x + i k_y y}, \qquad \qquad
\vert p \rangle = \frac{1}{q}\left( \begin{array}{c}
     q_x \\
     q_y 
\end{array} \right) e^{ik_x x + i k_y y}.
\end{split}
\label{S32}
\end{equation}

The fields of the mode propagating along the slab in the upper and lower media can be compactly written as the sum of the $s$- and $p$-polarized plane waves:

\begin{equation}
    \textbf{E}_1=\textbf{E}_1(x,y,z)=\sum_{\beta=s,p}a_{\beta}^1\vert \beta_{1} \rangle_+ e^{ik_z z}, \qquad \qquad  \textbf{E}_3=\textbf{E}_3(x,y,z)=\sum_{\beta=s,p}a_{\beta}^3\vert \beta_{3} \rangle_- e^{-ik_z z},
\label{S3Fields_isotr}
\end{equation}

\noindent with unknown amplitudes $a_{\beta}^{1,3}$. 

In contrast, the electric fields inside the biaxial slab ($0>z>-d$) should be represented with the help of the basis vectors found in Section \ref{sec:Frensel} as

\begin{equation}
    \textbf{E}_2=\textbf{E}_2(x,y,z)=\sum_{\gamma=o,e}a_{\gamma}^{2\downarrow}\vert \gamma \rangle_{+} e^{i k_{\gamma z} z} + a_{\gamma}^{2\uparrow} \vert \gamma \rangle_{-} e^{-i k_{\gamma z} z},
\label{S3Fields_biax}
\end{equation}

\noindent where $\vert \gamma\rangle_{\pm}$ denotes $\vert o \rangle_{\pm}$ and $\vert e \rangle_{\pm}$, being the polarization basis vector in the biaxial slab
\begin{equation}
\begin{split}
    \vert o \rangle_{\pm} = \frac{1}{q} \left(\begin{array}{c}
         -q_y (1 - \Delta_1 \Delta_z) \\
         q_x \\
         \mp iq_x q_y q_{oz} \Delta_1
    \end{array}  \right) e^{ik_x x + i k_y y},
    \qquad \qquad 
    \vert e \rangle_{\pm} = \frac{1}{q} \left(\begin{array}{c}
         q_x\frac{\Delta_2 - q_y^2}{\Delta_x^e} \\
         q_y \\
         \frac{\Delta_2}{\mp iq_{ez}}
    \end{array}  \right) e^{ik_x x + i k_y y}.
\end{split}
\label{Bas_vect_D}
\end{equation}
The factors $a_{\gamma}^{2\uparrow}$ and $a_{\gamma}^{2\downarrow}$ represent the unknown amplitudes of the plane waves travelling along and opposite to the $z$-axis, respectively.
Analogously to isotropic regions, we can introduce the in-plane subvectors $\vert o \rangle$ and $\vert e \rangle$ of the vectors $\vert o \rangle_{\pm}$, $\vert e \rangle_{\pm}$, respectively. These subvectors  can be compactly written as 
\begin{equation}
    \vert o \rangle = \vert s \rangle + \frac{q_yc_1}{q}\vert u_x \rangle,\qquad
    \vert e \rangle = \vert p \rangle + \frac{q_xc_2}{q}\vert u_x \rangle,
\label{2Doe}
\end{equation}
where $\vert u_x \rangle = (1,0)^T$, and 
\begin{equation}
    c_1 = \Delta_1 \Delta_z,\qquad    c_2 = \left(\frac{\Delta_2 - q_y^2}{\Delta_x^e} -1 \right).
\label{const_c}
\end{equation}

To find the magnetic fields we use Maxwell's equation, $\nabla \times \textbf{E}= -\frac{1}{c} \frac{\partial\textbf{H}}{\partial t}$. In case of a plane wave it simplifies to
\begin{equation}
    \textbf{H} = \textbf{q}_i \times \textbf{E}, 
\label{S3MagnF}
\end{equation}
\noindent where $\textbf{q}_i$ is the normalized wave vector of a corresponding plane wave. Then, we can apply the boundary conditions, which imply the continuity of the in-plane components of both electric and magnetic fields on the faces of the film (at $z=0$ and at $-d$):
\begin{equation}
\begin{split}
    \textbf{E}_{1t} (z=0)=\textbf{E}_{2t}(z=0), \qquad \qquad \qquad \textbf{H}_{1t}(z=0)=\textbf{H}_{2t}(z=0),\\
     \textbf{E}_{2t} (z=-d)=\textbf{E}_{3t}(z=-d), \qquad \qquad \ \textbf{H}_{2t}(z=-d)=\textbf{H}_{3t}(z=-d),
     \end{split}
    \label{S3BondCond}
\end{equation}

\noindent where the subscript "t" in Eqs. (\ref{S3BondCond}) stands for the in-plane subvectors. According to Eq. (\ref{S3MagnF}), the in-plane subvectors of the magnetic field can be written as $\textbf{H}_{t}=\textbf{e}_z\times  \textbf{q}_i \times \textbf{E}$.

Using the field representation (\ref{S3Fields_isotr}), (\ref{Bas_vect_D}), we can rewrite the boundary condition at $z=0$ in Eqs. (\ref{S3BondCond}) in a more explicit way:
\begin{equation}
    \sum_{\beta=s,p}a_{\beta}^{1}\vert \beta \rangle= \sum_{\gamma=e,o}a_{\gamma}^{2\downarrow}\vert \gamma \rangle + a_{\gamma}^{2\uparrow}\vert \gamma \rangle,
    \label{S3_Boudz=0E}
\end{equation}
\begin{equation}
    \textbf{e}_z\times \textbf{q}_{1+} \times \sum_{\beta=s,p}a_{\beta}^1 \vert \beta_1 \rangle_{+} = \textbf{e}_z\times\sum_{\gamma=e,o}\left(a_{\gamma}^{2\downarrow}\textbf{q}_{\gamma+} \times \vert \gamma \rangle_{+} + a_{\gamma}^{2\uparrow} \textbf{q}_{\gamma-} \times \vert \gamma \rangle_{-}\right),
    \label{S3_Boudz=0H}
\end{equation}

\noindent where $\textbf{q}_{1,3\pm}=\left(q_x, q_y, \pm i q_{1,3z}\right)^T$ and  $\textbf{q}_{\gamma\pm}=\left(q_x, q_y, \pm i q_{\gamma z}\right)^T$. To simplify Eq. (\ref{S3_Boudz=0H}), let us introduce auxiliary three-dimensional vectors, $\vert \beta_{1,3} \rangle'_{\pm}$ and $\vert \gamma\rangle'_{\pm}$:
\begin{equation}
\begin{split}
    \vert \beta_{1,3} \rangle'_{\pm} = - \textbf{e}_z\times \textbf{q}_{1,3\pm}\times \vert \beta_{1,3}\rangle_{\pm}, \\
    \vert \gamma\rangle'_{\pm} = - \textbf{e}_z\times \textbf{q}_{\gamma\pm} \times \vert \gamma\rangle_{\pm},
    \end{split}
    \label{S3_B'G'}
\end{equation}
where $\beta=s,p$ and  $\gamma = o,e$. Calculating the vector products in Eq. (\ref{S3_B'G'}) we obtain the following explicit relations for the in-plane two-dimensional subvectors $\vert s_{1,3} \rangle'$ and $\vert p_{1,3} \rangle'$:
\begin{equation}
\begin{split}
    \vert s_{1,3} \rangle' = - Y_s^{1,3} \vert s \rangle, \qquad \qquad
    \vert p_{1,3} \rangle' = - Y_p^{1,3} \vert p \rangle,
    \end{split}
    \label{S3_DefY}
\end{equation}
\noindent being $Y_\beta^{1,3}$ the admittances for the s-- and p-- polarized waves:
\begin{equation}
    Y_s^1 = iq_{1z}, \qquad \qquad Y_p^1 = \frac{\varepsilon_{1}}{iq_{1z}}, \qquad \qquad Y_s^3 = - i q_{3z}, \qquad \qquad Y_p^3 = - \frac{\varepsilon_{3}}{iq_{3z}}.
    \label{S3_Y}
\end{equation}
\noindent Since according to Eq. (\ref{S3_B'G'}), the $z$-component of the three-dimensional vectors $\vert \gamma\rangle'_{\pm}$ is 0, we keep the same notation for their two-dimensional in-plane subvectors: $\vert o\rangle'_{\pm}$ and $\vert e\rangle'_{\pm}$, which read explicitly as
\begin{align}
\vert o\rangle'_{\pm} = \pm i q_{oz} \left[ \vert s \rangle + q_y\Delta_1 \vert a \rangle \right], \qquad \qquad
\vert e\rangle'_{\pm} = \frac{\Delta_2}{\pm i q_{ez}}\vert p \rangle \pm iq_{ez} \vert b \rangle,
    \label{S3_O'E'}
\end{align}
\noindent with auxiliary vectors
\begin{equation}
    \vert a \rangle = \frac{1}{q} \left(\begin{array}{c}
         \Delta_z + q_x^2 \\
         q_x q_y
    \end{array}  \right), \qquad \qquad
        \vert b \rangle =\frac{1}{q} \left(\begin{array}{c}
         q_x (c_2+1) \\
         q_y
    \end{array}  \right).
    \label{S3_ab}
\end{equation}
As a result, using definition (\ref{S3_B'G'}) and Eqs. (\ref{S3_DefY}), we obtain a simple form of the Eq. (\ref{S3_Boudz=0H}):
\begin{equation}
    \sum_{\beta=s,p}a_{\beta}^1Y_\beta^1\vert \beta \rangle= \sum_{\gamma=o,e}\left(a_{\gamma}^{2\downarrow}\vert \gamma\rangle'_{+} + a_{\gamma}^{2\uparrow}\vert \gamma\rangle'_{-}\right).
    \label{S3_BoundSimpl}
\end{equation}
If we multiply (\ref{S3_Boudz=0E}) and (\ref{S3_BoundSimpl}) by $\langle\beta\vert$ [here, only the exponential of the bra-vector should be complex conjugated, for example, $\langle s\vert = (-q_y \; \; q_x)e^{-i k_x x - i k_y y}$] and taking into account that $\langle\beta\vert\beta'\rangle = \delta_{\beta, \beta'}$,  we get the following system of equations corresponding to the boundary condition at the interface $z=0$:
\begin{equation}
\begin{split}
     \sum_{\gamma=o,e}\left(a_{\gamma}^{2\downarrow}\langle\beta\vert\gamma\rangle   + a_{\gamma}^{2\uparrow}\langle\beta\vert\gamma\rangle \right) -a_{\beta}^{1}=0,\\
    \sum_{\gamma=o,e}\left(a_{\gamma}^{2\downarrow}\langle\beta\vert\gamma\rangle'_{+} + a_{\gamma}^{2\uparrow}\langle\beta\vert\gamma\rangle'_{-} \right) -a_{\beta}^1Y_\beta^1=0.
    \end{split}
    \label{S3_EqEHz=0}
\end{equation}

Analogously, for the interface $z=-d$, with the help of the auxiliary vectors $\vert\gamma\rangle'_{\pm}$ we find
\begin{equation}
\begin{split}
     \sum_{\gamma=o,e}\left(a_{\gamma}^{2\downarrow}\langle\beta \vert\gamma\rangle e^{q_{\gamma z} k_0 d}  + a_{\gamma}^{2\uparrow}\langle\beta\vert\gamma\rangle e^{-q_{\gamma z} k_0 d}\right) -a_{\beta}^{3}=0,\\
    \sum_{\gamma=o,e}\left(a_{\gamma}^{2\downarrow}\langle\beta\vert\gamma\rangle'_{+} e^{q_{\gamma z} k_0 d}+ a_{\gamma}^{2\uparrow}\langle\beta\vert\gamma\rangle'_{-} e^{-q_{\gamma z} k_0 d}\right) -a_{\beta}^3Y_\beta^3=0.
    \end{split}
    \label{S3_EqEHz=-d}
\end{equation}
Equations (\ref{S3_EqEHz=0}) and (\ref{S3_EqEHz=-d}) form a system of eight linear equations with eight unknowns. By defining $\xi^{\gamma\downarrow}=e^{q_{\gamma z} k_0 d}$ and $\xi^{\gamma\uparrow}=e^{-q_{\gamma z} k_0 d}$ with $\gamma=o,e$ (for the waves propagating along and opposite to z- axis, respectively), we have:

\begin{equation}
    \left(\begin{array}{cccccccc}
     -1 & 0 & \langle s \vert o \rangle & \langle s \vert o \rangle & \langle s \vert e \rangle & \langle s \vert e \rangle & 0 & 0\\
     0 & -1 & \langle p \vert o \rangle & \langle p \vert o \rangle & \langle p \vert e \rangle & \langle p \vert e \rangle & 0 & 0\\
     -Y_s^1 & 0 & \langle s \vert o \rangle'_{+} & \langle s \vert o \rangle'_{-} & \langle s \vert e \rangle'_{+} & \langle s \vert e \rangle'_{-} & 0 & 0\\
     0 & -Y_p^1 & \langle p \vert o \rangle'_{+} & \langle p \vert o \rangle'_{-} & \langle p \vert e \rangle'_{+} & \langle p \vert e \rangle'_{-} & 0 & 0\\
     0 & 0 & \langle s \vert o \rangle\xi^{o\downarrow} & \langle s \vert o \rangle\xi^{o\uparrow} & \langle s \vert e \rangle\xi^{e\downarrow} & \langle s \vert e \rangle\xi^{e\uparrow} & -1 & 0 \\
     0 & 0 & \langle p \vert o \rangle\xi^{o\downarrow} & \langle p \vert o \rangle\xi^{o\uparrow} & \langle p \vert e \rangle\xi^{e\downarrow} & \langle p \vert e \rangle\xi^{e\uparrow} & 0 & -1 \\
     0 & 0 & \langle s \vert o \rangle'_{+}\xi^{o\downarrow} & \langle s \vert o \rangle'_{-}\xi^{o\uparrow} & \langle s \vert e\rangle'_{+}\xi^{e\downarrow} & \langle s \vert e \rangle'_{-}\xi^{e\uparrow} & -Y_s^3 & 0 \\
     0 & 0 & \langle p \vert o \rangle'_{+}\xi^{o\downarrow} & \langle p \vert o \rangle'_{-}\xi^{o\uparrow} & \langle p \vert e \rangle'_{+}\xi^{e\downarrow} & \langle p \vert e \rangle'_{-}\xi^{e\uparrow} & 0 & -Y_p^3 \\
\end{array}
\right)\left(\begin{array}{c}
    a_s^1 \\ a_p^1 \\ a_o^{2\downarrow} \\ a_o^{2\uparrow} \\ a_e^{2\downarrow} \\ a_e^{2\uparrow} \\ a_s^3 \\ a_p^3
\end{array}
\right)=0.
 \label{S3_MatB}
\end{equation}

\noindent Using the explicit expressions for the vectors $\vert s\rangle$ and $\vert p\rangle$, $\vert o\rangle$ and $\vert e\rangle$, and $\vert o\rangle'_{\pm}$,  $\vert e\rangle'_{\pm}$, given by Eqs. (\ref{S32}), (\ref{2Doe}), and (\ref{S3_O'E'}), respectively, the scalar products in Eq. (\ref{S3_MatB}) can be explicitly calculated as
\begin{equation}
    \begin{split}
        & \langle s \vert o \rangle = \eta_1,\\
        & \langle p \vert o \rangle = \eta_2, \\
        & \langle s \vert e \rangle = \eta_3,\\
        & \langle p \vert e \rangle = \eta_4, \\
        & \langle s \vert o \rangle'_{\pm} = \pm iq_{oz} \eta_1,\\
        & \langle p \vert o \rangle'_{\pm} = \pm iq_{oz} \eta_2 \frac{\varepsilon_z}{\Delta_z},\\
        & \langle s \vert e \rangle'_{\pm} = \pm iq_{ez} \eta_3,\\
        & \langle p \vert e \rangle'_{\pm} = \frac{\Delta_2}{\pm i q_{ez}} \pm iq_{ez} \eta_4,
        \end{split}
    \label{S3_Terms_B}
\end{equation}
\noindent where $\eta_1 = 1-\frac{c_1 q_y^2}{q^2}$, $\eta_2 = \frac{q_xq_y\Delta_1}{q^2} (\Delta_z + q_x^2 + q_y^2)$, $\eta_3 = \frac{-q_xq_y c_2}{q^2}$, and $\eta_4 = 1 + \frac{q_x^2 c_2}{q^2}$.

The homogeneous system (\ref{S3_MatB}) has non-trivial solutions only when its determinant equals to zero. The zeros of the determinant yield the dispersion relation for the modes in the biaxial slab. In general, the dispersion relation can be analyzed numerically, but in the limit of a small slab thickness, $k_0d\ll 1$, as well as in the short-wavelength limit (large values of $q$),  it can be written in a compact analytical form, as will be shown below, in Sections  \ref{sec:Ultrathin} and \ref{sec:Large_q}, respectively. Before considering these interesting limits, we will ensure that our dispersion relation analytically reproduces some known examples.

\subsection{\label{sec:level2} Uniaxial slab}
Consider a uniaxial crystal with the axis $C$ pointing along the $z$-axis, $C\| Oz$, so that $\varepsilon_x= \varepsilon_y= \varepsilon_\bot$ and $\varepsilon_z= \varepsilon_\|$. In this case $\Delta_1=0$ and $\Delta_2= q_{ez}^2 + \varepsilon_\bot$, yielding $c_1 = c_2 = 0$, $\eta_2 = \eta _3 = 0$ and $\eta_1= \eta_4=1$. Then the scalar products given by Eq. (\ref{S3_Terms_B}) greatly simplify:
\begin{equation}
    \begin{split}
        & \langle s \vert o \rangle=\langle p \vert e \rangle  = 1,\\
        & \langle p \vert o \rangle=\langle s \vert e \rangle=\langle p \vert o \rangle'_{\pm}=\langle s \vert e \rangle'_{\pm} = 0, \\
                   & \langle s \vert o \rangle'_{\pm} = \pm i q_{oz},\\
        & \langle p \vert e \rangle'_{\pm} = \frac{\Delta_2}{\pm i q_{e z}} \pm i q_{e z} = \frac{\varepsilon_\bot}{\pm i q_{e z}}.
        \end{split}
    \label{S3_Terms_U}
\end{equation}

Consequently, the system of equations (\ref{S3_MatB}) reduces to

\begin{widetext}
\begin{equation}
\left(\begin{array}{cccccccc}
     -1 & 0 & 1 & 1 & 0 & 0 & 0 & 0\\
     -q_{1z} & 0 & q_{o z} & -q_{o z} & 0 & 0 & 0 & 0\\
     0 & -1 & \xi^{o\downarrow} & \xi^{o\uparrow} & 0 & 0 & 0 & 0 \\
     0 & q_{3z} & q_{o z}\xi^{o\downarrow} & -q_{o z}\xi^{o\uparrow} & 0 & 0 & 0 & 0 \\
     0 & 0 & 0 & 0 & -1 & 0 & 1 & 1 \\
     0 & 0 & 0 & 0 & -\frac{\varepsilon_1}{q_{z1}} & 0 & \frac{\Delta_2}{q_{e z}} & -\frac{\Delta_2}{q_{e z}}\\
     0 & 0 & 0 & 0 & 0 & -1 & \xi^{e\downarrow} & \xi^{e\uparrow} \\
     0 & 0 & 0 & 0 & 0 & \frac{\varepsilon_3}{q_{3z}} & \frac{\Delta_2}{q_{e z}}\xi^{e\downarrow} & -\frac{\Delta_2}{q_{e z}}\xi^{e\uparrow} \\
\end{array}
\right)
\left(\begin{array}{c}
    a_s^1 \\ a_s^3 \\ a_o^{2\downarrow} \\ a_o^{2\uparrow} \\ a_p^1 \\ a_p^3 \\ a_e^{2\downarrow} \\ a_e^{2\uparrow}
\end{array}
\right)=0.
 \label{S3_MatU}
\end{equation}
\end{widetext}

Vanishing of the determinant of the matrix in Eq. (\ref{S3_MatU}) results in two separate equations, yielding (after some straightforward algebra) the dispersion of the ordinary and extraordinary modes:
\begin{equation}
    \text{ordinary:} \quad \text{tanh}(q_{o z} k_0 d)=-\frac{q_{oz} (q_{1z} + q_{3z})}{q_{1z} q_{3z} + q_{o z}^2},
    \label{S3_Uo}
\end{equation}
\begin{equation}
    \text{extraordinary:} \quad \text{tanh}(q_{e z} k_0 d)=-\frac{q_{e z} \varepsilon_\bot(q_{1z}\varepsilon_3+q_{3z} \varepsilon_1)}{q_{1z} q_{3z} \varepsilon_\bot^2 + q_{e z}^2 \varepsilon_1 \varepsilon_3}.
    \label{S3_Ue}
\end{equation}

\subsection{\label{sec:level2} Isotropic slab}

The dispersion of the modes in an isotropic slab can be easily derived from the dispersion of the modes in the uniaxial slab, by setting $\varepsilon_\bot= \varepsilon_\parallel = \varepsilon_2$. By doing so, the $z$-components of the wave vectors $q_{o z}$, $q_{e z}$ degenerate to $q_{2z}^2 =q^2 - \varepsilon_2$ and Eqs. (\ref{S3_Uo}), (\ref{S3_Ue}) transform into the well-known expressions for electromagnetic TE and TM modes in a conventional slab waveguide, respectively:
\begin{equation}
\text{TE:} \quad \text{tanh}(q_{2z} k_0 d)=-\frac{q_{2z}(q_{1z} + q_{3z})}{q_{1z}q_{3z} + q_{2z}^2},
\label{S3_IsTE}
\end{equation}
\begin{equation}
\text{TM:} \quad \text{tanh}(q_{2z} k_0 d)=-\frac{q_{2z} \varepsilon_2 (q_{1z}\varepsilon_3 + q_{3z}\varepsilon_1)}{q_{1z} q_{3z}\varepsilon_2^2 + q_{2z}^2 \varepsilon_1 \varepsilon_3}.
\label{S3_IsTM}
\end{equation}

\section{\label{sec:Thick}Very thick slabs: surface modes at the biaxial crystal boundaries}

Let us consider now another extreme case, assuming that the thickness of the slab tends to infinity, $d \rightarrow \infty$. Then our dispersion relation should split into two independent dispersion relations describing surface modes at the interfaces between the biaxial crystal and two isotropic media with
dielectric permittivities $\varepsilon_1$ and $\varepsilon_3$. To obtain these dispersion relations in a simple analytical form, we multiply the third and fifth columns of the determinant of the system \eqref{S3_MatB} by $\xi^{o\uparrow}$  and  $\xi^{e\uparrow}$, respectively. Then tending $d \rightarrow \infty$ in the determinant, and assuming that both $q_{oz}$ and $q_{ez}$ have a non-vanishing real part, we see that all the matrix elements proportional to $\xi^{o,e\uparrow}$ (third and fifth elements in the four first rows and fourth and
sixth ones in the four last rows) vanish. As a result, the $8\times 8$ determinant becomes a product of the two determinants $4\times 4$, each of them describing the surface modes at the 1-2 ($z=0$) and 2-3 ($z=-d$) interfaces. Without loss of generality, let us consider only one of these determinants $4\times 4$, corresponding to the interface 1-2. Zeroing the determinant, we have
\begin{equation}
    \begin{vmatrix}
     -1 & 0 & \langle s \vert o \rangle & \langle s \vert e \rangle\\
     0 & -1 & \langle p \vert o \rangle & \langle p \vert e \rangle\\
     -Y_s^1 & 0 & \langle s \vert o' \rangle_{-} & \langle s \vert e' \rangle_{-}\\
     0 & -Y_p^1 & \langle p \vert o' \rangle_{-} & \langle p \vert e' \rangle_{-}\\
\end{vmatrix}=0.
 \label{S5_MatS}
\end{equation}

Then, using the Gauss method, we can reduce the dimension of the matrix to $2\times 2$, as
\begin{equation}
\begin{vmatrix}
Y_s^1 \langle s \vert o \rangle - \langle s \vert o' \rangle_{-} & Y_s^1 \langle s \vert e \rangle - \langle s \vert e' \rangle_{-} \\
Y_p^1 \langle p \vert o \rangle - \langle p \vert o' \rangle_{-} & Y_p^1 \langle p \vert e \rangle - \langle p \vert e' \rangle_{-} \\
\end{vmatrix} = 0.
\label{S5Det}
\end{equation}

To write the dispersion relation in a compact form, we express $\varepsilon_z$ from Frensel's equation for biaxial slabs \eqref{S2FrEq} as a function of $q_{oz}$:
\begin{equation}
\varepsilon_z = \frac{\varepsilon_x \varepsilon_y q^2 + \left(q_{oz}^2 - q^2\right)\left(\varepsilon_x q_x^2 + \varepsilon_y q_y^2\right)}{q_{oz}^2 \left(\varepsilon_x + \varepsilon_y + q_{oz}^2 - q^2\right) + \varepsilon_x \varepsilon_y - \varepsilon_x q_x^2 - \varepsilon_y q_y^2}.
\label{S5Ez}
\end{equation}

\noindent Then, using the identities $q^2=\varepsilon_1 + q_{1z}^2$ and $q^2=q_{x}^2 + q_{y}^2$, we substitute $\varepsilon_z$ from Eq. (\ref{S5Ez}) into the expressions for $Y_{s,p}^1$ in Eq. \eqref{S3_Y} and scalar products given by Eq. \eqref{S3_Terms_B}, appearing in the elements of the matrix in Eq. \eqref{S5Det}. After some algebraic operations, Eq. \eqref{S5Det} reproduces the dispersion relation for surface waves on boundaries of biaxial crystals, derived in Ref. \cite{Narimanov18}:
\begin{equation}
\left(q_{1z}+q_{o z}\right) \left(q_{1z}+q_{e z}\right) \left(\varepsilon_x \varepsilon_y - \varepsilon_x q_x^2 - \varepsilon_y q_y^2 - \varepsilon_1 q_{o z} q_{e z}\right) - q_{o z} q_{e z} \left(\varepsilon_1 - \varepsilon_x \right) \left(\varepsilon_1 - \varepsilon_y\right)=0.
\label{S5Disp}
\end{equation}

\subsection{\label{sec:level2}Uniaxial crystal with the axis perpendicular to the interface}
Consider a uniaxial crystal with the axis $C$ along the $z$-axis, thus directed perpendicularly to the interface of the crystal. Defining, as before,  $\varepsilon_x= \varepsilon_y= \varepsilon_\bot$ and $\varepsilon_z= \varepsilon_\|$, and taking into account that according to Eq. (\ref{oe}), $\varepsilon_\bot =q^2 - q_{o z}^2$, Eq. \eqref{S5Disp} simplifies as:
\begin{equation}
q_{o z} \left(q_{1z}+q_{o z}\right) \left(q_{1z} + q_{e z}\right) \left(\varepsilon_\bot q_{o z} + \varepsilon_1 q_{e z}\right) + q_{o z} q_{e z} \left(q_{1z} + q_{o z}\right)^2 \left(q_{1z} - q_{o z}\right)^2=0.
\label{S5DispU1}
\end{equation}

\noindent Since $\text{Re}(q_{o z})>0$ and $\text{Re}(q_{1z})>0$, and therefore $q_{o z} \left(q_z + q_{o z}\right) \neq 0$, we can divide Eq. \eqref{S5DispU1} by $q_{o z} \left(q_z + q_{o z}\right)$. Then it transforms to:
\begin{equation}
q_{e z} \varepsilon_1 \left(q_{o z} + q_{e z}\right) + q_{1z} \varepsilon_\bot \left(q_{o z} + q_{e z}\right) = 0.
\label{S5DispU2}
\end{equation}

\noindent Assuming that $q_{e z} + q_{o z} \neq 0$, we obtain the dispersion relation for the surface wave on a boundary of a uniaxial crystal:
\begin{equation}
\frac{q_{e z}}{\varepsilon_\bot} + \frac{q_{1z}}{\varepsilon_1} = 0,
\label{S5DispUf}
\end{equation}

\noindent where $q_{1z}=\sqrt{q^2 - \varepsilon_1}$ and $q_{ez}=\sqrt{\frac{\varepsilon_\bot}{\varepsilon_\parallel} q^2 - \varepsilon_\bot}$. Deriving $q$ from this equation, the dispersion relation takes the well-known form (see e.g. Ref. \cite{Agranovich82}):

\begin{equation}
q = \sqrt{\varepsilon_1 \varepsilon_\parallel \frac{\varepsilon_1 - \varepsilon_\bot}{\varepsilon_1^2 - \varepsilon_\parallel \varepsilon_\bot}}.
\label{S5DispUfAM}
\end{equation}

In the isotropic case, $\varepsilon_\parallel=\varepsilon_\bot = \varepsilon$, Eq. \eqref{S5DispUf} simplifies to the dispersion relation for surface waves at the interface between
two isotropic media:
\begin{equation}
q = \sqrt{\frac{\varepsilon_1 \varepsilon}{\varepsilon_1 + \varepsilon}}.
\label{S5DispUI}
\end{equation}

\subsection{\label{sec:level2}Uniaxial case with in-plane crystal axis}

Consider a uniaxial crystal with the $C$ axis along the $y$-axis, thus lying in the plane of the interface of the crystal. Redefining here $\varepsilon_x = \varepsilon_z= \varepsilon_\bot$ and $\varepsilon_y= \varepsilon_\parallel$, Eq. \eqref{S2WV} for $q_{o,ez}$ transforms to (similar to Eq. (\ref{oe})):
\begin{equation}
q_{oz}^{2}=q^{2} - \varepsilon_\bot, \qquad \qquad
\varepsilon_\bot q_{ez}^{2}=\varepsilon_\bot q_x^{2} - \varepsilon_\parallel q_y^2 - \varepsilon_\bot \varepsilon_\parallel.
\label{S5_qoe}
\end{equation}
Substituting the expression for $q_{ez}$ from Eq. \eqref{S5_qoe} into Eq. \eqref{S5Disp} and dividing it by $q_{ez}$, we get the famous dispersion relation for Dyakonov surface waves \cite{Dyakonov88}:

\begin{equation}
\left(q_{1z} + q_{o z}\right) \left(q_{1z} + q_{e z}\right) \left(\varepsilon_1 q_{o z} + \varepsilon_\bot q_{e z}\right) + q_{o z} \left(\varepsilon_1 - \varepsilon_\bot \right) \left(\varepsilon_1 - \varepsilon_\parallel\right)=0.
\label{S5_DispDyak}
\end{equation}

\section{\label{sec:Ultrathin}Ultrathin slab limit}

Recently, polaritons in ultra-thin slabs and monolayers (for instance, plasmon polaritons in a monolayer graphene \cite{Goncalves16} or hyperbolic phonon polaritons in thin slabs of polar dielectrics, such as $h$-BN \cite{Caldwell19}) have attracted particularly high attention. Therefore, the limit of a vanishing slab thickness $d \rightarrow 0$ is of a great practical interest. Let us illustrate how our general dispersion relation, given by the determinat of the system (\ref{S3_MatB}), can be simplified for ultra-thin slabs. Analogously to the methodology used for isotropic slabs \cite{Nikitin17}, we can approximate the slab of a finite thickness by a two-dimensional conductive sheet, with the effective conductivity, $\hat{\sigma}$, given by $\hat{\sigma} =\frac{\omega d \hat{\varepsilon}}{4\pi i}$. To that end, let us assume that all the components of the tensor $\hat{\varepsilon}$ are large, i.e. $|\varepsilon_i| \gg 1$ $(i=x,y,z)$.

\begin{figure}[H]
\centering
\includegraphics[width=8.5cm]{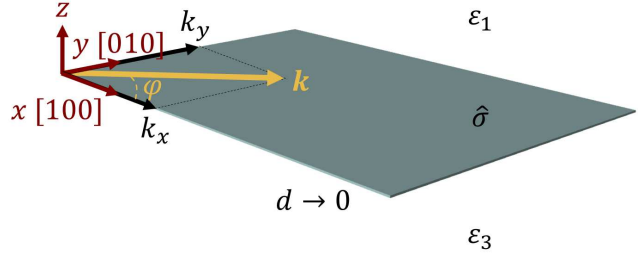}
\caption{\label{Fig2} Schematics of the ultra-thin crystal slab.}
\label{SNOM}
\end{figure}
Then, retaining in Eq. (\ref{S2WV}) the first non-vanishing terms depending upon $q_x$ and $q_y$ in the expressions for the normalized $z$-components of the wave vectors, $q_{o,e z}$, Eq. (\ref{S2WV}) can be greatly simplified:

\begin{equation}
q_{o,e z}^{2}=\frac{1}{2}\left\lbrace -(\varepsilon_{x}+\varepsilon_{y})+ \frac{\varepsilon_{x}+\varepsilon_{z}}{\varepsilon_{z}}q_{x}^{2} +\frac{\varepsilon_{y} +\varepsilon_{z}}{\varepsilon_{z}} q_{y}^{2} \right\rbrace
\pm \frac{1}{2} \left\lbrace\varepsilon_{x} -\varepsilon_{y} +\frac{\varepsilon_{z} -\varepsilon_{x}}{\varepsilon_{z}} q_{x}^{2} -\frac{\varepsilon_{z} - \varepsilon_{y}}{\varepsilon_{z}} q_{y}^{2}\right\rbrace,
\label{S4WV}
\end{equation}
where  "$+$" and "$-$" should be taken for the labels "$o$", and "$e$", respectively. Equations (\ref{S4WV}) then further simplify to
\begin{equation}
\begin{split}
q_{o z}^{2}=-\varepsilon_{y} + q_{x}^{2} + \frac{\varepsilon_{y}}{\varepsilon_{z}} q_{y}^{2},\\
q_{e z}^{2}=-\varepsilon_{x} + q_{y}^{2} + \frac{\varepsilon_{x}}{\varepsilon_{z}} q_{x}^{2}. 
\end{split}
\label{S4WVOE}
\end{equation}
\noindent Using Eq. (\ref{S4WVOE}), we find $ c_1=-\frac{\varepsilon_{z}}{q_{y}^{2}}$ and  $c_2=-1$, so that the scalar products  (\ref{S3_Terms_B}) can be written as
\begin{equation}
\begin{split}
& \langle s \vert o \rangle = \frac{q_x^2}{q^2},\\
& \langle p \vert o \rangle = \frac{q_x q_y}{q^2}, \\
& \langle s \vert e \rangle = \varepsilon_z \frac{\varepsilon_x - \varepsilon_y}{\varepsilon_x - \varepsilon_z} \frac{q_y}{q_x q^2},\\
& \langle p \vert e \rangle = - \varepsilon_z \frac{\varepsilon_x - \varepsilon_y}{\varepsilon_x - \varepsilon_z} \frac{1}{q^2}, \\
& \langle s \vert o \rangle'_{\pm} = \pm iq_{oz} \langle s \vert o \rangle,\\
& \langle p \vert o \rangle'_{\pm} = \pm iq_{oz} \langle p \vert o \rangle,\\
& \langle s \vert e \rangle'_{\pm} = \pm iq_{ez} \langle s \vert e \rangle,\\
& \langle p \vert e \rangle'_{\pm} = \pm iq_{ez} \langle p \vert e \rangle.
\end{split}
\label{S4MCoef}
\end{equation}

Additionally, assuming the small thickness of the slab, we can simplify the elements of the matrix in Eq. (\ref{S3_MatB}) by expanding the exponentials $\xi^{\gamma}$ into the Taylor series in $k_0d$ and retaining the first non-vanishing terms. We have $\xi^{\gamma\downarrow}=e^{q_{\gamma z} k_0 d}=1+q_{\gamma z} k_0 d$ and $\xi^{\gamma\uparrow}=e^{-q_{\gamma z} k_0 d}=1-q_{\gamma z} k_0 d$. To simplify the determinant of the system (\ref{S3_MatB}), we sum up its third and fifth columns with the fourth and sixth columns, respectively, and then subtract the fourth and sixth columns (both multiplied by the factor $1/2$), from the third and fifth columns, respectively. Then using row operations, we eliminate two first and two last columns in the obtained determinant by the Gauss method: we multiply first (second) row to $Y_s^1$ ($Y_p^1$) and subtract it from the third (fourth) row and analogously for fifth (sixth) and seventh (eighth) rows. As a result, we get the following equation:
\begin{equation}
\begin{vmatrix}
i \langle s \vert o \rangle & -Y_s^1 \langle s \vert o \rangle & i \langle s \vert e \rangle & -Y_s^1 \langle s \vert e \rangle\\
i \langle p \vert o \rangle & -Y_p^1 \langle p \vert o \rangle & i \langle p \vert e \rangle & -Y_p^1 \langle p \vert e \rangle\\
-Y_s^3 k_0 d \langle s \vert o \rangle & \left(2\alpha_y + Y_s^1 - Y_s^3\right) \langle s \vert o \rangle & -Y_s^3 k_0 d \langle s \vert e \rangle & \left(2\alpha_x + Y_s^1  -Y_s^3\right) \langle s \vert e \rangle \\
-Y_p^3 k_0 d \langle p \vert o \rangle & \left(2\alpha_y + Y_p^1 - Y_p^3\right) \langle p \vert o \rangle & -Y_p^3 k_0 d \langle p \vert e \rangle & \left(2\alpha_x + Y_p^1 - Y_p^3\right) \langle p \vert e \rangle \\
\end{vmatrix} = 0,
\label{S4Mat2}
\end{equation}

\noindent where $\alpha_{x,y}= \frac{2\pi \sigma_{x,y}}{c}=\frac{k_0 d\varepsilon_{x,y}}{2i }$ are the normalized 2D effective conductivity components. Using the smallness of $k_0 d$, on the one hand and the assumed large values of the components of the tensor $\hat{\varepsilon}$, on the other hand, the determinant (\ref{S4Mat2}) can be further significantly simplified. Namely, a more detailed analysis (which we omit here) shows that  the elements proportional to $k_0 d$ (the first and third elements of the third and fourth rows) yield the contribution of a higher order of smallness and thus can be neglected. As a result, the determinant (\ref{S4Mat2}) factorizes into a product of the two determinants of the sub-matrices $2\times 2$:

\begin{equation}
\begin{vmatrix}
\langle s \vert o \rangle & \langle s \vert e \rangle \\
\langle p \vert o \rangle & \langle p \vert e \rangle
\end{vmatrix}
\cdot
\begin{vmatrix}
\left(2\alpha_y + Y_s^1 - Y_s^3\right) q_x & \left(2\alpha_x + Y_s^1  -Y_s^3\right) q_y \\
\left(2\alpha_y + Y_p^1 - Y_p^3\right) q_y & -\left(2\alpha_x + Y_p^1 - Y_p^3\right) q_x \\
\end{vmatrix} = 0.
\label{S4Mat3}
\end{equation}

Taking into account that the first determinant in Eq. \eqref{S4Mat3} gives $-\frac{\varepsilon_z}{q^2} \frac{\varepsilon_x - \varepsilon_y}{\varepsilon_x - \varepsilon_z} \neq 0$, the dispersion relation is given by the vanishing of the the second determinant:

\begin{equation}
\left\lbrace \alpha_x q_{y}^{2} + \alpha_y q_{x}^{2} + \frac{q_{x}^{2} + q_{y}^{2}}{2} \left(i q_{1z}+i q_{3z}\right) \right\rbrace \left\lbrace \alpha_x q_{x}^{2} + \alpha_y q_{y}^{2} + \frac{q_{x}^{2} + q_{y}^{2}}{2} \left(\frac{\varepsilon_{1}} {i q_{1z}} + \frac{\varepsilon_{3}}{i q_{3z}}\right) \right\rbrace=q_{x}^{2} q_{y}^{2}\left(\alpha_x - \alpha_y\right)^2.
\label{S4Disp2D}
\end{equation}

\noindent This dispersion relation, written for biaxial slabs of a small but nonzero thickness (the effective conductivities $\alpha_{x,y}$ are thickness-dependent), has been used for the analysis of hyperbolic phonon polaritons in thin slabs of $\alpha$-MoO\textsubscript{3} \cite{Ma18}. Nevertheless, to our knowledge, it has not been consistently derived for a nonvanishing slab thickness up to now. For 2D anisotropic sheets (of zero thickness) Eq. (\ref{S4Disp2D}) is exact and had been reported in Refs. \cite{GomezDiaz15, Yermakov15}.
As expected, the asymptotes of the dispersion relation (\ref{S4Disp2D}) ($q_{x,y}\rightarrow\infty$) coincide with the ones, following from the Fresnel equations [compare with Eq. (\ref{asF})]:
\begin{equation}
\frac{q_x}{q_y} =\sqrt{-\frac{\alpha_y}{\alpha_x}}= \sqrt{-\frac{\varepsilon_y}{\varepsilon_x}}.
\label{as2D}
\end{equation}
In case of an isotropic 2D sheet $\alpha_x=\alpha_y=\alpha$ and Eq. (\ref{S4Disp2D}) splits into two independent equations describing the dispersion of the TE and TM modes \cite{Hanson08} propagating along the sheet:
\begin{equation}
\begin{split}
     &\text{TE}: \quad q_{1z}+q_{3z}-2i\alpha=0,\\
     &\text{TM}: \quad \frac{\varepsilon_1}{q_{1z}}+\frac{\varepsilon_3}{q_{3z}} + 2i\alpha=0.
\end{split}
\label{S4Graph}
\end{equation}

\begin{figure}[H]
\centering
\includegraphics[width=16cm]{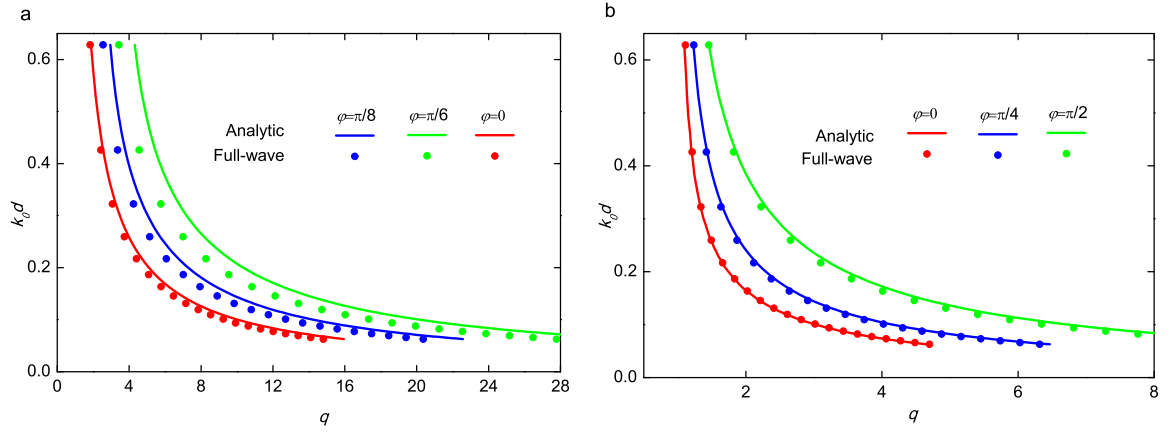}
\caption{\label{Fig3} Comparison of the dispersion relation given by  Eq. \eqref{S4Disp2D} and by the full-wave numerical simulations (\textsc{COMSOL MULTIPHYSICS}). In both panels $\varepsilon_1=\varepsilon_3=1$. Permittivity tensor components: \textbf{a)} $\varepsilon_x = -2$, $\varepsilon_y = 2$, $\varepsilon_z = 2$. \textbf{b)} $\varepsilon_x = -7$, $\varepsilon_y = -3$, $\varepsilon_z = 2$.}
\label{SNOM}
\end{figure}

To demonstrate the validity of the simplified dispersion relation (\ref{S4Disp2D}), we compare in Fig. \ref{Fig3} the refractive indices of a mode found from (\ref{S4Disp2D}) (solid curves) to those found from full-wave simulations (points). Figures \ref{Fig3}(a) and \ref{Fig3}(b) represent the result for two different illustrative sets of parameters: in Fig. \ref{Fig3}(a) only one of the in-plane dielectric permittivities is negative, $\varepsilon_x<0$ (while $\varepsilon_y,\varepsilon_z>0$), and in Fig. \ref{Fig3}(b) both in-plane permittivities are negative, $\varepsilon_x,\varepsilon_y<0$ (while $\varepsilon_z>0$). Both in Figs. \ref{Fig3}(a) and \ref{Fig3}(b) the propagation of the mode at different angles $\varphi$ (see Fig. \ref{Fig2}) is considered. As expected, the agreement between the analytical approximation and rigorous numeric simulations improves for smaller values of $k_0d$, although in Fig. \ref{Fig3}(b) the agreement is good in the whole shown range of $k_0d$. Impressively, the agreement between the analytical and numerical results is in general excellent for all the shown propagation directions,
even though neither $k_0d$ is very small nor the values of $\varepsilon_{x}$, $\varepsilon_{y}$, and  $\varepsilon_{z}$ are very large, as it was initially assumed for the derivation of (\ref{S4Disp2D}).

\section{\label{sec:Large_q}The limit of a large refractive index of the modes}

The general dispersion relation given by the zeroing of the determinant in Eq. (\ref{S3_MatB}) can be also greatly simplified under the assumption of large refractive indices of the modes, $\vert q \vert \gg 1$. Such simplification is similar to the one made for the dispersion of the modes in uniaxial crystal slabs \cite{Dai14}.  For large $q$ the expressions for the $z$-components of the wave vectors inside the slab can be approximated as:

\begin{equation}
q_{o z}^{2}= q^2 - \frac{\varepsilon_x \varepsilon_z q_y^2 + \varepsilon_y \varepsilon_z q_x^2-\varepsilon_x\varepsilon_y q^2}{\varepsilon_z q^2 - \varepsilon_x q_x^2 - \varepsilon_y q_y^2}, \qquad \qquad q_{e z}^{2}=\frac{\varepsilon_x}{\varepsilon_z} q_x^2 + \frac{\varepsilon_y}{\varepsilon_z} q_y^2,
\label{S6WVO}
\end{equation}
where we have retained the second-order (the highest-order) term in $q$ both in $q_{e z}^2$ and $q_{o z}^2$, as well as the zeroth-order term in the expression for $q_{o z}^{2}$, to avoid uncertainty in $\Delta_1$ (since $\Delta_x^o=q_x^2$ and $\Delta_z=q_{o z}^2=q^2$, we have $\Delta_1=\frac{0}{0}$). Substituting the Eqs. \eqref{S6WVO} into Eqs. (\ref{const_delta}), (\ref{const_c}), we have

\begin{equation}
\begin{split}
& \Delta_1=\frac{1}{q_{o z}^2} \frac{\varepsilon_x -\varepsilon_y}{\varepsilon_z- \varepsilon_x},\quad \Delta_2=q_{e z}^2, \\
& c_1=\frac{\varepsilon_x -\varepsilon_y}{\varepsilon_x- \varepsilon_z},\quad c_2=\frac{\varepsilon_y -\varepsilon_x}{q_{e z}^2- q_{o z}^2},
\end{split}
\label{S6_DelC}
\end{equation}

\noindent where we have neglected all small amendments in the expressions for each constant. Then we obtain the simplified expressions for the scalar products (\ref{S3_Terms_B}):

\begin{equation}
\begin{split}
& \langle s \vert o \rangle = \frac{\varepsilon_z}{\varepsilon_x-\varepsilon_z} \frac{q_{e z}^2- q_{o z}^2}{q_{o z}^2},\\
& \langle p \vert o \rangle = \frac{q_x q_y}{q_{o z}^2} \frac{\varepsilon_x - \varepsilon_y}{\varepsilon_x - \varepsilon_z},\\
& \langle s \vert e \rangle = \frac{q_x q_y}{q_{o z}^2} \frac{\varepsilon_x- \varepsilon_y}{q_{e z}^2 - q_{o z}^2}, \\
& \langle p \vert e \rangle = 1,\\
& \langle s \vert o' \rangle_{\pm} = \pm i q_{o z} \langle s \vert o \rangle, \\
& \langle p \vert o' \rangle_{\pm} = \pm i\varepsilon_z\frac{q_x q_y}{q_{o z}^3}  \frac{\varepsilon_x - \varepsilon_y}{\varepsilon_z - \varepsilon_x},\\
& \langle s \vert e' \rangle_{\pm} = \pm i q_{e z} \langle s \vert e \rangle ,\\
& \langle p \vert e' \rangle_{\pm} =  \mp i q_{e z} \frac{\varepsilon_z}{q_{o z}^2} .
\end{split}
\label{S6_Coef}
\end{equation}

To simplify the matrix \eqref{S3_MatB} and to eliminate the two first and two last columns, we use the same column and row operations as in the Section \ref{sec:Ultrathin} and obtain the following equation (containing the $4\times 4$ matrix):

\begin{equation}
{\begin{vmatrix}
\langle s \vert o' \rangle_{+} & -Y_s^1 \langle s \vert o \rangle & \langle s \vert e' \rangle_{+} & -Y_s^1 \langle s \vert e \rangle \\
\langle p \vert o' \rangle_{+} & -Y_p^1 \langle p \vert o \rangle & \langle p \vert e' \rangle_{+} & -Y_p^1 \langle p \vert e \rangle \\
\langle s \vert o' \rangle_{+} C_o - Y_s^3 \langle s \vert o \rangle S_o & \langle s \vert o' \rangle_{+} S_o - Y_s^3 \langle s \vert o \rangle C_o & \langle s \vert e' \rangle_{+} C_e - Y_s^3 \langle s \vert e \rangle S_e & \langle s \vert e' \rangle_{+} S_e - Y_s^3 \langle s \vert e \rangle C_e \\
\langle p \vert o' \rangle_{+} C_o - Y_p^3 \langle p \vert o \rangle S_o& \langle p \vert o' \rangle_{+} S_o - Y_p^3 \langle p \vert o \rangle C_o & \langle p \vert e' \rangle_{+} C_e - Y_p^3 \langle p \vert e \rangle S_e & \langle p \vert e' \rangle_{+} S_e - Y_p^3 \langle p \vert e \rangle C_e
\end{vmatrix}} = 0,
\label{S6Mat1}
\end{equation}

\noindent where, for compactness, we have introduced abbreviated notations for the hyperbolic functions:

\begin{equation}
   C_{o,e} = \text{cosh}(q_{o,e z} k_0 d), \quad  S_{o,e} = \text{sinh}(q_{o,e z} k_0 d).
    \label{cossin}
\end{equation}

Let us notice that the matrix elements containing either $\langle s \vert o' \rangle_{+}$ or $Y_s^{1,3} \langle s \vert o \rangle$ are of order $\sim q$ , while the other matrix elements are of order $\sim \frac{1}{q}$, thus being much smaller in magnitude. Consequently, we can neglect the third and fourth elements of the first and third rows (the contribution of these elements to the determinant is of the second and fourth order of smallness in $\frac{1}{q}$). As a result, the determinant in Eq. \eqref{S6Mat1} factorizes into a product of two determinants of the sub-matrices $2\times 2$, so that Eq. \eqref{S6Mat1} splits into the two following equations:

\begin{equation}
\begin{split}
\begin{vmatrix}
\langle s \vert o' \rangle_{+} & -Y_s^1 \langle s \vert o \rangle \\
\langle s \vert o' \rangle_{+} C_o - Y_s^3 \langle s \vert o \rangle S_o & \langle s \vert o' \rangle_{+}S_o - Y_s^3 \langle s \vert o \rangle C_o \\
\end{vmatrix} = 0, \\
\\
\begin{vmatrix}
\langle p \vert e' \rangle_{+} & -Y_p^1 \langle p \vert e \rangle \\
\langle p \vert e' \rangle_{+} C_e - Y_p^3 \langle p \vert e \rangle S_e & \langle p \vert e' \rangle_{+}S_e - Y_p^3 \langle p \vert e \rangle C_e \\
\end{vmatrix} = 0.
\end{split}
\label{S6Mat2}
\end{equation}

Simplifying all the admittances as $Y_p^1 = \frac{\varepsilon_1}{q_{1z}} \approx \frac{\varepsilon_1}{q} \approx \frac{\varepsilon_1}{q_{oz}}$, $Y_p^3 = -\frac{\varepsilon_3}{q_{3z}} \approx -\frac{\varepsilon_3}{q} \approx -\frac{\varepsilon_3}{q_{oz}}$ and $Y_s^1 \approx -Y_s^3 \approx i q \approx i q_{oz}$, we can easily calculate both determinants \eqref{S6Mat2}. Vanishing of the first determinant does not give any physically reasonable solutions since

\begin{equation}
    -\frac{2 \varepsilon_z^2 \left(q_{ez}^2 - q_{oz}^2\right)^2 e^{q_{oz} k_0 d}}{\left(\varepsilon_x - \varepsilon_z\right)^2 q_{oz}^2} \neq 0.
    \label{S61stDet}
\end{equation}

Therefore, the dispersion relation follows from the vanishing of the second determinant in Eq. \eqref{S6Mat2}:

\begin{equation}
\text{tanh}(q_{e z} k_0 d)=-\frac{\left(\varepsilon_1 + \varepsilon_3\right) \varepsilon_z q_{o z} q_{e z}}{\varepsilon_1 \varepsilon_3 q_{o z}^2 + \varepsilon_z^2 q_{e z}^2}.
\label{S6DispRel}
\end{equation}

To write Eq. \eqref{S6DispRel} in a convenient form, let us define
\begin{equation}
\rho=i\sqrt{\frac{\varepsilon_zq^2}{\varepsilon_x q_x^2 +\varepsilon_y q_y^2}}=i\sqrt{\frac{\varepsilon_z}{\varepsilon_x \text{cos}^2\varphi +\varepsilon_y \text{sin}^2\varphi}},
\label{S6rho}
\end{equation}
where $\varphi$ is the angle between the $x$ axis and the in-plane component of wave vector. Then using Eq. \eqref{S6WVO} (neglecting here the second term in $q_{oz}^2$), Eq. \eqref{S6DispRel} can be written as

\begin{equation}
 \text{tan}\left(\frac{q k_0 d}{\rho}\right) = \frac{\rho \frac{\varepsilon_1 + \varepsilon_3}{\varepsilon_z}}{1 - \frac{\rho^2 \varepsilon_1 \varepsilon_3}{\varepsilon_z^2}}.
\label{S6DispSimpl}
\end{equation}

Taking into account that $\text{arctan}\left( \frac{x+y}{1-xy}\right) = \text{arctan}(x) + \text{arctan}(y)$, we get a simple expression for the normalized in-plane wave vector $q$ in the biaxial slab in the short-wavelength limit, $q\gg 1$:

\begin{equation}
q=\frac{\rho}{k_0 d}\left[\text{arctan} \left( \frac{\varepsilon_1 \rho}{\varepsilon_z} \right) + \text{arctan}\left(\frac{\varepsilon_3 \rho}{\varepsilon_z} \right)+\pi l\right], \qquad l=0,1,2... 
\label{S6q}
\end{equation}
We can verify that Eq. (\ref{S6q}) transforms into the dispersion of modes in a uniaxial slab (with the axis $C$ along the $z$-axis),  setting $\varepsilon_x = \varepsilon_y = \varepsilon_\bot$ and $\varepsilon_z = \varepsilon_\parallel$. Then defining $\psi=-\rho = -i \sqrt{\frac{\varepsilon_\parallel}{\varepsilon_\bot}}$, and taking into account that $\frac{\varepsilon_{1,3} \rho}{\varepsilon_\parallel} = \frac{\varepsilon_{1,3}}{\psi \varepsilon_\bot}$, we reproduce the dispersion relation, used for the analysis of hyperbolic phonon polaritons in $h$-BN crystal slabs \cite{Dai14}:
\begin{equation}
q=-\frac{\psi}{k_0 d}\left[\text{arctan} \left( \frac{\varepsilon_1}{\psi \varepsilon_\bot}\right) + \text{arctan}\left(\frac{\varepsilon_3}{\psi \varepsilon_\bot} \right)+\pi l\right], \qquad l=0,1,2... 
\label{S6qUn}
\end{equation}
The same results can be straightforwardly derived from the exact Eq. (\ref{S3_Ue}), in the limit of large $q$. Equation \eqref{S6q} also reduces to Eq. \eqref{S6qUn} when the propagation of the mode coincides either with the $x$ axis (in this case we should set  $\varepsilon_x = \varepsilon_\bot$ and  $\varepsilon_z = \varepsilon_\parallel$) or with the $y$-axis  (in this case we should set  $\varepsilon_y = \varepsilon_\bot$ and  $\varepsilon_z = \varepsilon_\parallel$). In the two latter particular cases, anisotropic polaritons in $\alpha$-MoO\textsubscript{3} slabs were studied via Eq. \eqref{S6qUn} in Ref. \cite{Zheng18}.

To verify the validity of our analytical approximation, we compare the isofrequency curves obtained from Eq. \eqref{S6q} with those obtained from full-wave electromagnetic simulations. As an example, we take the slab thickness $d = 100$ nm and the free-space wavelength $\lambda=1\; \mu $m. We consider the following four different combinations of the (purely real) permittivity tensor components:
\begin{equation}
    \begin{split}
        \text{a) } \varepsilon_x<0,\; \varepsilon_y<0,\; \varepsilon_z>0,\\
        \text{b) } \varepsilon_x<0,\; \varepsilon_y<0,\; \varepsilon_z<0,\\
        \text{c) } \varepsilon_x<0,\; \varepsilon_y>0,\; \varepsilon_z>0,\\
        \text{d) } \varepsilon_x<0,\; \varepsilon_y>0,\; \varepsilon_z<0.\\
    \end{split}
    \label{S7Eps}
\end{equation}

\begin{figure}[H]
\centering
\includegraphics[width=16cm]{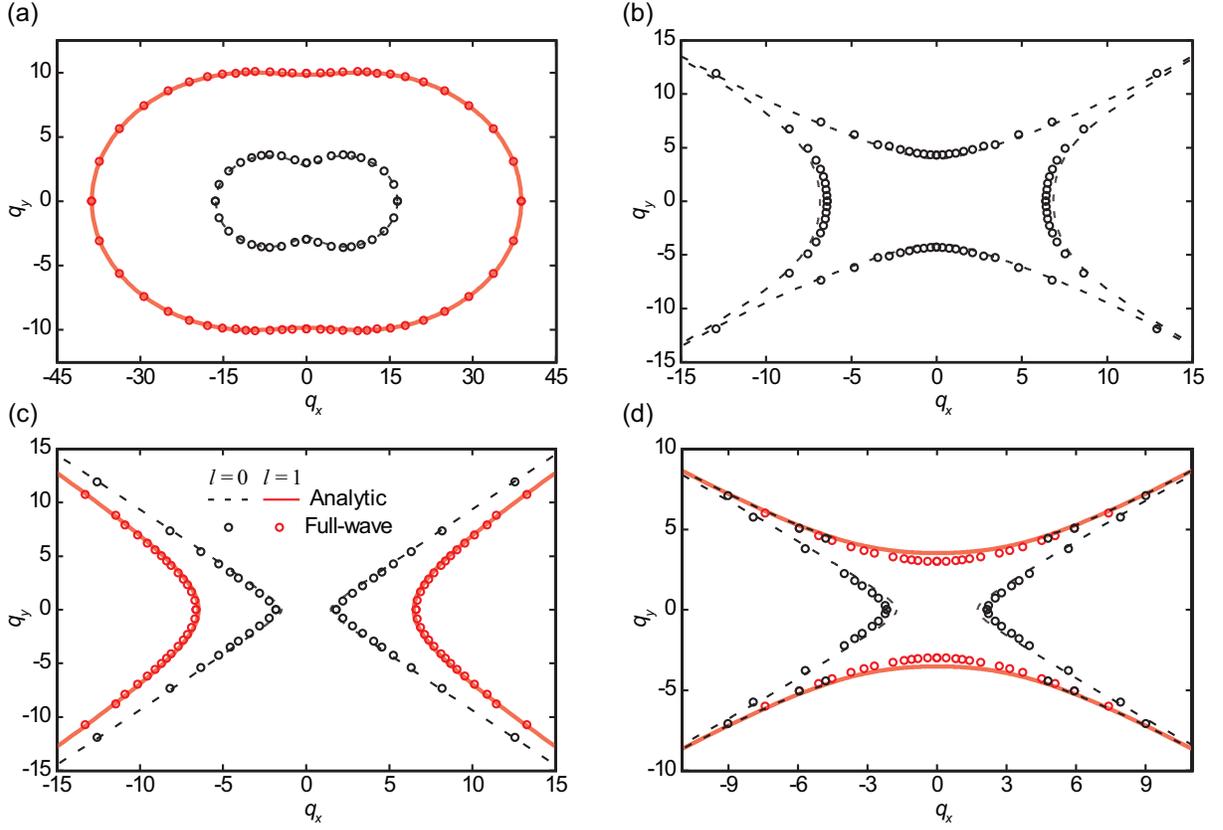}
\caption{\label{Fig4} Comparison between the isofrequency curves of the modes in a biaxial crystal slab. The isofrequency curves found according to Eq. \eqref{S6q} are plotted by the discontinuous black curves ($l=0$) and continuous red curves ($l=1$), while those found from the full-wave simulations are plotted by the black and red dots, respectively. In all panels $\lambda=1$ $\mu$m, $d = 100$ nm and $\varepsilon_1=\varepsilon_3=1$. Permittivity tensor components: \textbf{a)} $\varepsilon_x = -0.1$, $\varepsilon_y = -1$, $\varepsilon_z = 2$. \textbf{b)} $\varepsilon_x = -0.1$, $\varepsilon_y = -1$, $\varepsilon_z = -2$. \textbf{c)} $\varepsilon_x = -2$, $\varepsilon_y = 2$, $\varepsilon_z = 2$. \textbf{d)} $\varepsilon_x = -2$, $\varepsilon_y = 2$, $\varepsilon_z = -2$.}
\label{SNOM}
\end{figure}

\noindent Fig. \ref{Fig4} shows the isofrequency curves extracted from both Eq. \eqref{S6q} (red and black curves) and full-wave numeric simulations (dots). For the parametric sets a), c), d) corresponding to the volume modes, the isofrequency curves for the two lowest modes are shown: $l=0$ (black discontinuous curve and black dots) and $l=1$ (red continuous curve and red dots). In contrast, for the set of parameters b), the mode exponentially decays inside and outside the slab (so that it has a surface wave character) and therefore only the solution with $l=0$ makes sense. In all panels of Fig. \ref{Fig4}  we see an excellent agreement between the numeric simulations and analytical approximations for large $q$ ($q\gtrsim10$) and even very reasonable agreement for $q$ comparable to 1.  This agreement, particularly for the case of small and moderate values of $q$, unambiguously evidences that Eq. (\ref{S6q}) can be used in a wide space of parameters for the characterization of diverse modes in both natural and artificial biaxial crystal slabs. 

\section{\label{sec:Conclusion}Conclusions}
To summarize, we have presented an analytical derivation of the electromagnetic modes that can be guided along biaxial crystal slabs. We have provided simple expressions for the dispersion of the modes in the limit of an ultra-thin slab and for the case of large $k$-vectors of the modes. Both limits are currently of great importance for studying highly-confined anisotropic polaritons in vdW biaxial crystal slabs and, particularly, for the interpretation of the state-of-the-art near-field experiments. 

\section{\label{sec:Acknoledgements}Acknowledgements}
We thank A. Bylinkin for checking the analytical
derivations. A.Y.N. acknowledges the Spanish Ministry
of Science, Innovation and Universities (national Project No. MAT2017-88358-C3-3-R) and Basque Government
(Grant No. IT1164-19). P.A.-G. acknowledges support from the European Research Council under Starting Grant No. 715496, 2DNANOPTICA. K.V.V. and V.S.V. acknowledge
support from the Russian Science Foundation, Grant No.
18-19-00684.

G.\'A-P. and K.V.V. contributed equally to this work.

\end{document}